\documentclass[useAMS,usenatbib,usegraphicsx]{mn2e}

\usepackage{graphicx} 
\usepackage{amssymb}  

\title[The Tully--Fisher relation of cluster spirals at $z$ = 0.83]
      {The Tully--Fisher relation of cluster spirals at $z$ = 0.83\thanks{%
       Based on observations collected at the European Southern Observatory,
       Chile (66.A--0376).}\thanks{%
       Based on observations with the NASA/ESA Hubble Space Telescope,
       obtained at the Space Telescope Science Institute, which is operated
       by AURA, Inc., under NASA contract NAS5-26555.}}

\author[B. Milvang-Jensen et al.]
  {Bo~Milvang-Jensen,$^1$\thanks{E-mail: ppxbm@nottingham.ac.uk},
   Alfonso~Arag{\'o}n-Salamanca,$^1$
   George~K.~T.~Hau,$^2$
   \newauthor
   Inger~J{\o}rgensen,$^3$
   Jens~Hjorth$^4$ \\
   $^1$School of Physics and Astronomy,
       University of Nottingham,
       University Park,
       Nottingham NG7 2RD, UK \\
   $^2$European Southern Observatory,
       Casilla 19001,
       Vitacura,
       Santiago, Chile\\
   $^3$Gemini Observatory,
       670 N. A'ohoku Place,
       Hilo,
       HI 96720, USA \\
   $^4$Astronomical Observatory,
       University of Copenhagen,
       Juliane Maries Vej 30,
       2100 Copenhagen {\O}, Denmark
}

\date{27--Nov--2002, accepted for publication in MNRAS}

\pagerange{\pageref{firstpage}--\pageref{lastpage}} \pubyear{2002}

\def\MS1054{{MS1054$-$03}}
\def\Vrotsini{{V_{\rm rot} \sin i}}
\def\Vrot    {{V_{\rm rot}}}
\def\slitangle{{\theta_{\rm slit}}}

\def\rdspec{{r_{\rm d,spec}}}
\def\rdphot{{r_{\rm d,phot}}}
\def\dTF{{\Delta{\rm TF}}}
\def\dTFcorr{{\Delta{\rm TF_{\rm corr}}}}

\def\Pdiffdistr{{P^{\textrm{\scriptsize K--S}}_{\rm diff.\;distr.}}}

\begin{document}

\label{firstpage}

\maketitle

\begin{abstract}
We present the rest-frame $B$--band Tully--Fisher relation for a sample
of 8 cluster spiral galaxies at $z=0.83$ and 19 field spirals at 
$z$ = 0.15--0.90 based on VLT spectroscopy and HST photometry. 
No strong difference is detected between the cluster and the field galaxies, 
but we find some evidence that 
the cluster spirals are $\sim0.5$--$1\,$mag brighter than
the field ones at a fixed rotation velocity. Although only a 
$\sim 1.5$--$2\sigma$ result, if confirmed with larger samples this 
effect could be due to the cluster spirals
experiencing a period of enhanced star formation while falling into the 
cluster. 
\end{abstract}

\begin{keywords}
galaxies: spiral --
galaxies: evolution --
galaxies: clusters: individual: MS1054.4$-$0321
\end{keywords}

\section{Introduction}

Ground-based and HST observations indicate that the disk galaxy population in
rich galaxy clusters has experienced remarkable evolution since $z = 1$. It has
been argued that the increase with time of the S0 fraction and the simultaneous
decrease in the spiral fraction suggest that star-forming spirals fall into
distant
clusters at a much higher rate than in the local Universe,
and that these spirals ultimately
become S0s when star formation is extinguished by the cluster environment
(e.g.\ \citealt{Dressler_etal:1997,Poggianti_etal:1999};
but see also \citealt{Andreon:1998}).  
Recent hydro-dynamical simulations of the interaction of the gaseous components
of disk galaxies with the intracluster medium support these ideas
\citep*{Quilis_etal:2000}.
They also indicate that while the gas is being removed from the disk
a brief period of enhanced star formation could be expected.
The strong evolution of the cluster spiral population contrasts with
the mild evolution observed in the field spirals to $z\sim 1$
\citep[][and references therein]{Vogt:1999:mn_quick_fix,Vogt:2001a:mn_quick_fix}.
To quantify the evolution of the cluster
spirals, we are conducting a programme 
measuring the stellar and dynamical masses and $M/L$ ratios
for a sizeable sample of morphologically-classified disk galaxies in rich
galaxy clusters at $0.2 \! < \! z \! < \! 0.9$.  We here present the first
results on the Tully--Fisher relation \citep{Tully_Fisher:1977}
of cluster galaxies at $z=0.83$.

We assume $H_0 = 75\,$ km$\,$s$^{-1}\,$Mpc$^{-1}$ and $q_0 = 0.05$
in order to easily compare with Vogt and collaborators.

\section{The data}

\subsection{Sample selection}

\begin{figure*}
\includegraphics[width=172mm]{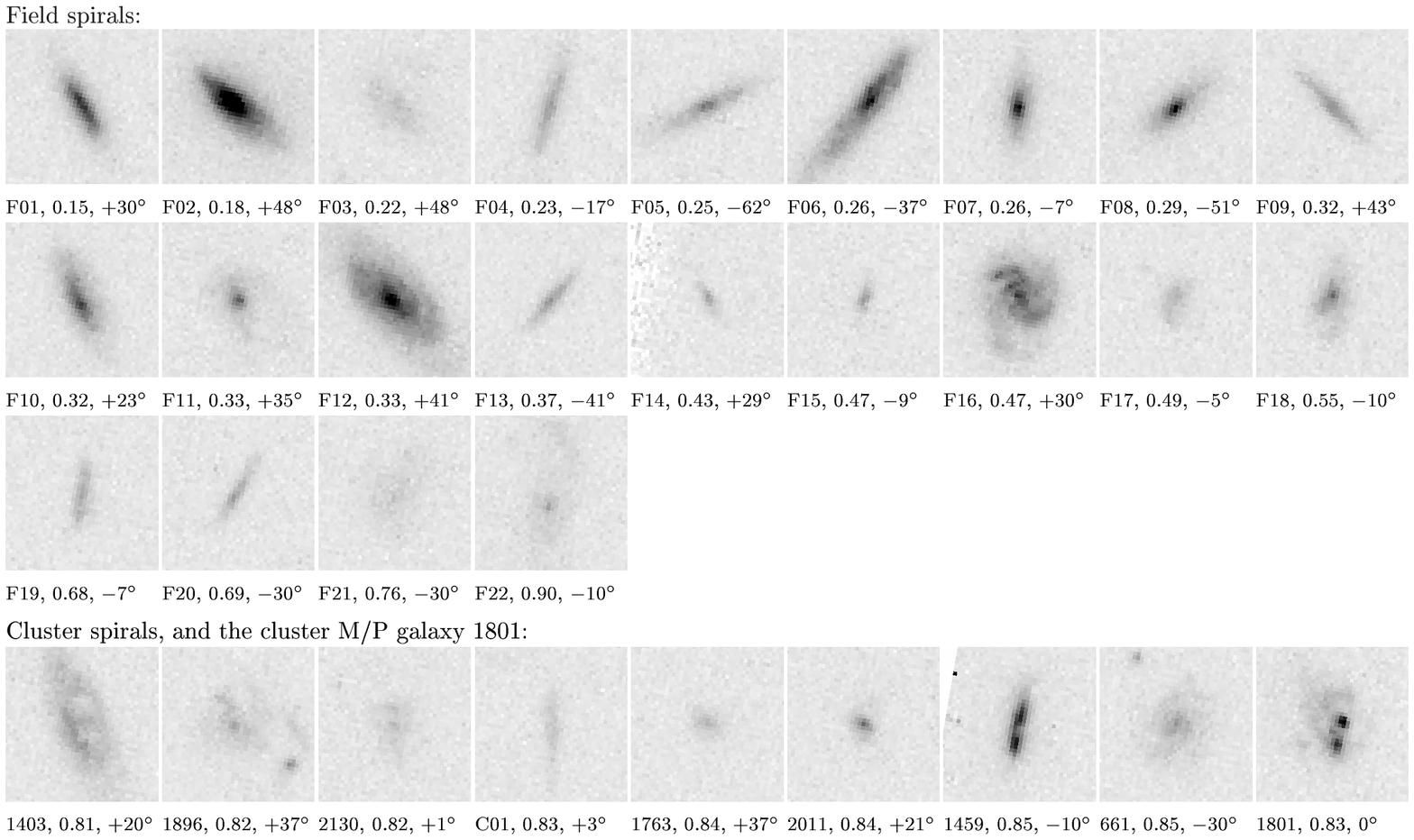}
\caption{HST+WFPC2 F814W images of the galaxies with detected emission.
Below each image is given galaxy ID, redshift and
slit angle ($\slitangle$, measured counter-clockwise from vertical).
The IDs follow \citet{vanDokkum:1999} for the cluster galaxies, with the
exception of C01 (a cluster galaxy found by us). 
The VLT slits were aligned with the major axes of the galaxies.
The images shown have been rotated to the mask position angle
and are 4$''$ on the side.
The intensity scaling is linear, and the intensity cuts are the same
for all the galaxies.
}
\label{fig:HST_montage}
\end{figure*}

We targeted the galaxy cluster {\MS1054} at $z=0.83$ since it is a rich  X-ray
selected cluster at high redshift and has extensive HST imaging and 8-m
spectroscopy available. We based our galaxy selection on 
the spectroscopic and photometric catalogues of known cluster members
of \citet{vanDokkum:1999} \citep[see also][]{vanDokkum_etal:2000}.
The spectroscopic catalogue was based on an $I$--band selected sample
with $20.0 < I < 22.7$ (corresponding to $-22.3 < M_B < -19.6$ at 
this redshift).  
The catalogue gives Keck-based H$\delta$ and [OII] equivalent widths (EWs) and
spectral types (Emission, Absorption or E+A). 
The photometric catalogue gives HST based magnitudes, colours and morphologies.
The combined catalogue contains 87 galaxies,
for which 74 have both
photometric and spectroscopic information.
 From this sample, we selected spiral galaxies as follows:
\begin{enumerate}
\item
Galaxies having spiral morphology and Emission spectral type (EW([OII])
$\ge 5\,${\AA}).  We included galaxy 2011, which has 
spiral morphology, although it was originally classified as merger/peculiar 
(M/P). 
Two galaxies  were removed due to geometric constraints (1733
and 1888X). 
This gave 6 galaxies, of which 5 yielded rotation velocities.
\item
Galaxies having spiral morphology and
a less secure Emission spectral type (EW([OII]) $\ge 5\,${\AA}, but with
larger errors). 
This gave 4 galaxies, but only 2 rotation velocities.
\item
Galaxies having spiral morphology and no listed spectral type.
After removing one galaxy (1354) due to geometry, 
this gave 2 Sa galaxies.
However, we were unable to measure rotation velocities for them.
\end{enumerate}
Twelve spiral galaxies in the catalogue (50\%) did not have
Emission spectral type, and these were not selected.
All galaxies were required to have an inclination of
$i > 30^\circ$ (with $0^\circ$ being face-on).
Finally, we included an M/P cluster galaxy 
with strong [OII] emission (1801) out of curiosity,
since it did not conflict with any of the high priority targets.
Thus, we started with a sample of 13 known cluster galaxies.
To fill the remaining space in our two spectroscopic masks, 
additional galaxies were selected. 
We morphologically classified the galaxies in the WFPC2 mosaic
which were not in the cluster catalogue of \citet{vanDokkum:1999}
and then selected galaxies with spiral/disk morphology
and $i>30^\circ$.
We imposed magnitude and colour limits corresponding to those spanned
by the selected cluster galaxies, i.e., 
F814W = 20.1--22.7 and (F606W$-$F814W) = 0.8--2.2.
When there was a geometrical conflict, the brightest galaxy 
was chosen.
In order to fill all gaps in the mask we sometimes
relaxed the magnitude and colour limits. 
There were relatively few galaxies in the red end of
our search window.
This supplementary sample of 34 spirals
yielded one additional cluster galaxy (C01, F814W = 22.25)
for which the rotation velocity 
was measured, and
19 $z$ = 0.15--0.90 field galaxies with rotation velocities.
The magnitude range for these was F814W = 19.1--23.2.
The field galaxies observed 
provided an ideal comparison sample since it was 
observed under the same conditions 
as our cluster sample.
A montage of WFPC2 images of the galaxies with detected emission is shown
in Fig.~\ref{fig:HST_montage}.

\begin{figure*}
\includegraphics[bb = 46 369 543 733,width=169mm]{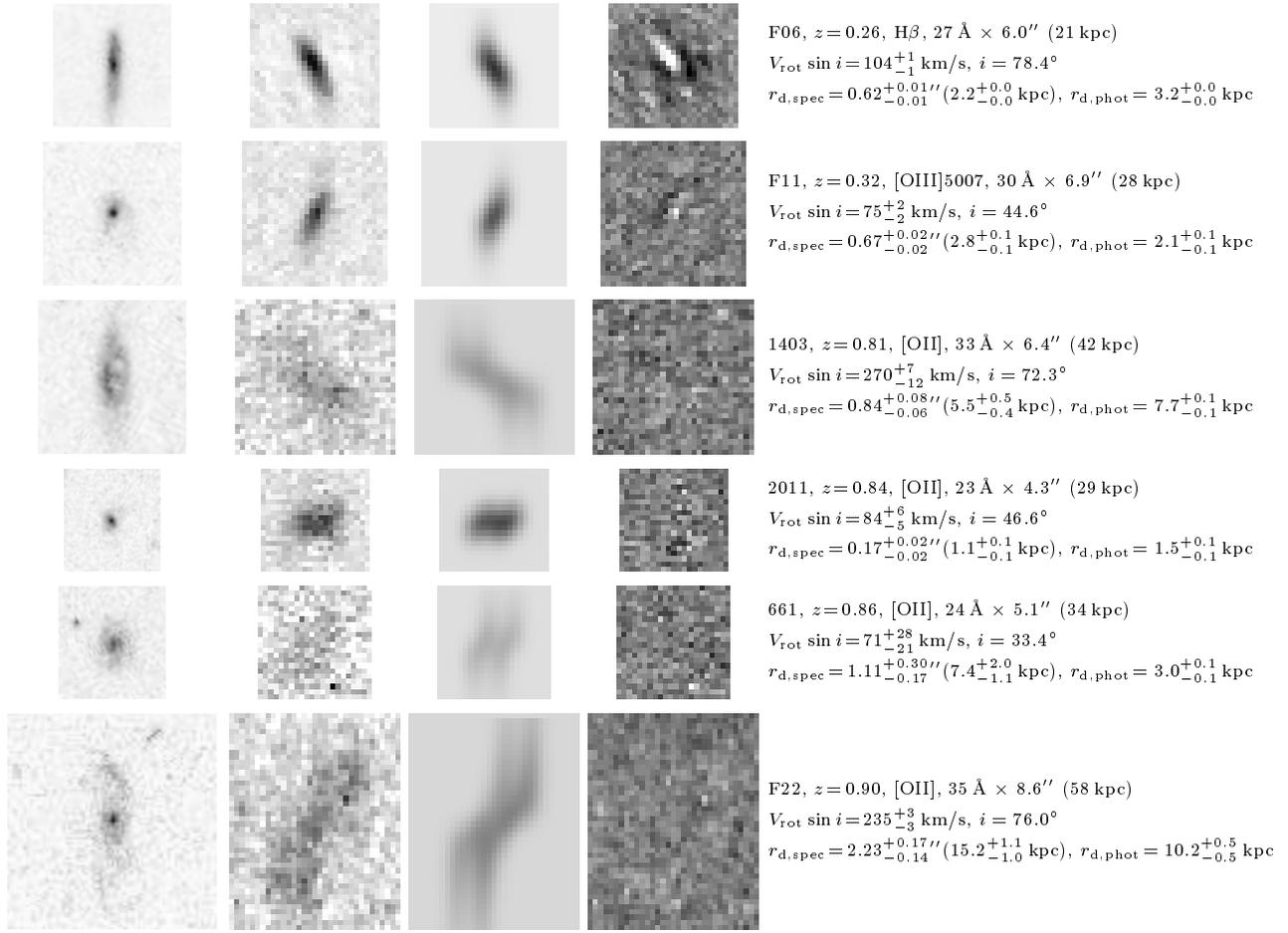}
\caption{Illustration of the emission line fitting.
The first column shows HST+WFPC2 F814W images of the 6 example galaxies,
rotated to have the slit along the $y$-axis.
The following columns show 2D spectral images:
observed, best-fit model and residual, with wavelength along the $x$--axis.
The intensity cuts have been adjusted from galaxy to galaxy,
except for the residual images.
The galaxy ID, redshift, line ID and
spectral image size are given on the figure,
as well as the fitted values of the projected rotation velocity ($\Vrotsini$)
and the emission line exponential scale length ($\rdspec$).
The inclination ($i$) and the F814W photometric scale length ($\rdphot$)
are also given.
For each galaxy the height in arcsec of all 4 images is the same.
We rejected the fit of galaxy F06, cf.\ Sect.~\ref{sec:ELFIT2D}.
}
\label{fig:ELFIT2D}
\end{figure*}

\subsection{Spectroscopy}
\label{sec:spec}

Observations were carried out with the FORS2 instrument at the
VLT (UT2) on 2001 Feb 23. 
Two spectroscopic masks were designed with position angles (PAs) 
at right angles to each other, in order to be able to cover all galaxy PAs.
The slits were aligned with the major axes of the galaxies,
tilting the slits with respect to the PA of the mask
by an angle $\slitangle$ (cf.\ Fig.~\ref{fig:HST_montage}).
The slits were 1$''$ wide in the dispersion direction.
Typical slit lengths were 11$''$.
The spectral resolution was FWHM = 4.2$\,${\AA}, and 
the pixel size 1.075$\,${\AA} $\times$ 0.201$''$.
The median wavelength range was 5400--7600$\,${\AA}. 
Each mask was observed for $7 \times 30\,$min.
The seeing in the combined frames
was 1.04$''$ for mask 1, and 0.94$''$ for mask 2.

Details of the spectroscopic data reduction will be given in  Milvang-Jensen
(2002, PhD Thesis, in prep.). Briefly,  after bias subtraction,  cosmic rays
were removed using the 7 frames taken   for each mask.  The geometric
distortion was mapped and removed using the edges of the spectra in the `dome'
flats.  Dome and sky flats were used to correct  for pixel-to-pixel variations
and    slit profiles respectively.  The science frames were then cut up into 
individual slit spectra, wavelength calibrated, and sky subtracted.

Depending on redshift, different emission lines were 
visually identified in the 2D spectra   
([OII]$\lambda\lambda$3726.1,3728.8, H$\gamma$, H$\beta$,
[OIII]$\lambda$4959, [OIII]$\lambda$5007 and H$\alpha$).

\subsection{Rotation velocities}
\label{sec:ELFIT2D}

Projected rotation velocities
were determined from the 2D emission line spectra
using the synthetic rotation curve method of
\citet{Simard_Pritchet:1998,Simard_Pritchet:1999}.
A simple parametrized intrinsic rotation curve  
and an exponential emission intensity profile are assumed.
Synthetic 2D emission line spectra are then produced taking
into account known observational parameters such as  
seeing, spectral resolution and slit-width.
The model parameters are:
projected rotation velocity $\Vrotsini$;
exponential scale length $\rdspec$;
total line intensity $I$;
residual background level $b$; and
[OII] doublet intensity ratio $R$. 
The main inputs are 
a continuum-subtracted postage stamp spectral image
of the emission line, an estimate of the background noise,
and the inclination $i$ (cf.\ Sect.~\ref{sec:Photometry}).
The Metropolis algorithm \citep{Metropolis_etal:1953} is used to search the
parameter space
to get `best fit' values and 68\% confidence intervals.

\begin{figure*}
\includegraphics[width=175mm]{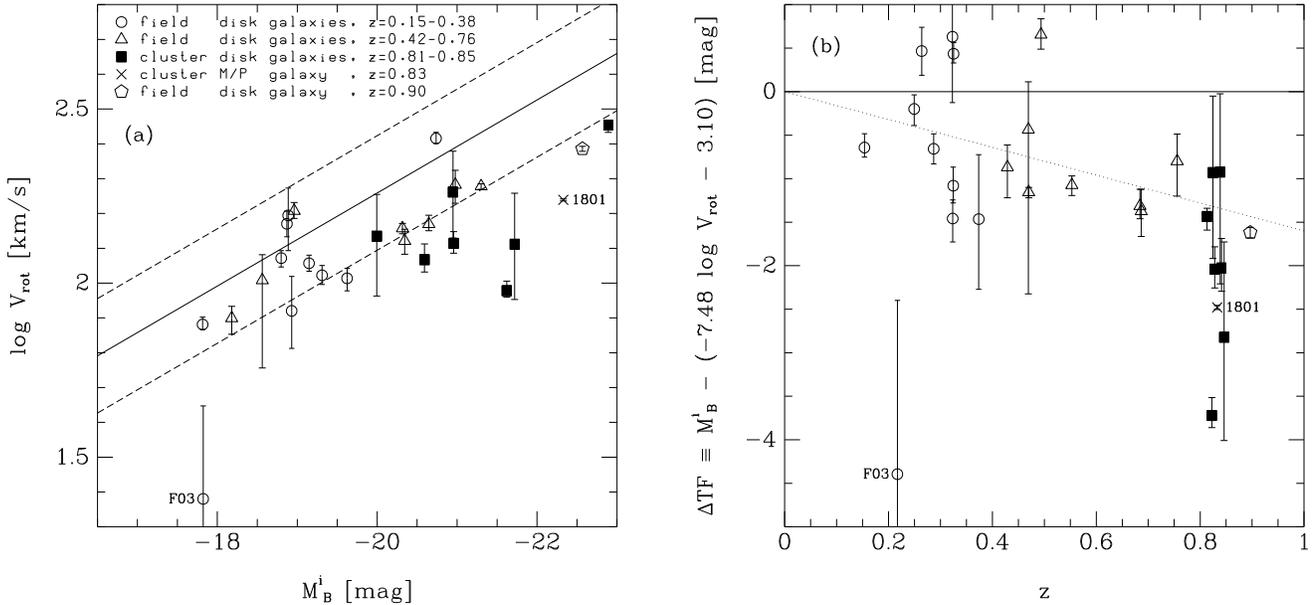}
\caption{(a)~High redshift cluster and field $B$--band Tully--Fisher relation.
Error bars on $M^i_B$ are not shown since they are smaller than the plot
symbols.
The solid line is the local
Tully--Fisher relation
from \citet{Pierce_Tully:1992}, cf.\ \citet{Vogt:1999:mn_quick_fix}:
$M^i_B = -7.48 \log\Vrot - 3.10$.
The dashed lines mark the $3\sigma$ limits ($\sigma = 0.41\,$mag).
Galaxy 1459 is off the scale of the plot at
$(M^i_B,\log\Vrot) = (-22.70,0.99)$.
(b)~Residuals from the local Tully--Fisher relation versus redshift.
The dotted line is a fit to the field galaxies
(except galaxy F03): $\dTF = (-1.6 \pm 0.3) z$.
}
\label{fig:Tully-Fisher}
\end{figure*}

The fitting was done for all well-detected emission lines.
Examples of emission lines, model spectra and residual images are
shown in Fig.~\ref{fig:ELFIT2D}. 
Out of the 31 galaxies with well-detected emission lines,
3 field galaxies did not yield acceptable fits. 
One (F04)
had a strange line morphology: the line showed a tilt only on one side
of the nucleus, whereas on the other the line was flat.
Two (F02 and F06)
had an intensity profile that looked more extended than
exponential. F06 is shown in Fig.~\ref{fig:ELFIT2D}.
The galaxies with rejected fits have been excluded from our analysis.
The M/P cluster galaxy (1801) was 
surprisingly well matched to the model. However,  
the emission was asymmetric, being strongest in the two bright knots 
(cf.\ Fig.~\ref{fig:HST_montage}). 
We show this galaxy in  Fig.~\ref{fig:Tully-Fisher}, but we do not include it
in the analysis.

Two different intrinsic rotation curves were tried:
a step-wise flat rotation curve and the `Universal' rotation curve of
\citet{Persic_Salucci:1991}.
The latter is a function of the luminosity of the galaxy.
The differences in the model images are small
due to the effect of the seeing and the relatively coarse sampling.
For the low luminosity galaxies the Universal rotation curve gives marginally
higher rotation velocities that the flat one. For the high luminosity galaxies
(i.e.\ the cluster galaxies and some of the field galaxies) there is little
difference between the results from two rotation curves. In the analysis we
will use the results from the Universal rotation curve, since this rotation
curve has some physical motivation.
For galaxies with more than one emission line in the observed wavelength range, 
the rotation velocities derived from the different lines agreed 
within the errors,
and weighted averages were used in the analysis.  

The ratio $\rdspec/\rdphot$
had a median value of 1.1 and a typical range of 0.7--1.7,
in broad agreement with what is found locally \citep{Ryder_Dopita:1994}.
This indicates that the fitting procedure is producing sensible results.
The fitted $\rdspec$ was typically in the range 0.2--1.2$''$,
and almost always greater than zero at the 3$\sigma$ level.
Spectral signal is detectable and thus contributes to the fit
up to typically $\gtrsim$ 2 scale lengths.

\subsection{Photometry}
\label{sec:Photometry}

Photometry was carried out on the F814W and F606W  HST+WFPC2 images 
\citep{vanDokkum_etal:1999}.  The typical exposure time for the combined HST
images was 3.3$\,$ksec.  
Photometric zero points were taken from
the May 1997 WFPC2 SYNPHOT update.
The SExtractor programme \citep{Bertin_Arnouts:1996}
was used to measure total magnitudes. 
Rest-frame $B$ magnitudes
were derived from the observed F814W and F606W photometry. F814W matches
almost exactly the rest-frame $B$--band at $z=0.83$, and thus the $B$
magnitudes can be derived very accurately for galaxies 
close to this $z$. 
F606W matches the $B$-band at $z \approx 0.37$. Thus,  $B$ magnitudes 
for all the galaxies in our sample can be interpolated with reasonably small
uncertainties. The interpolation (and in a few cases, small extrapolation) was
carried out using spectral energy distributions (SEDs) of local galaxies with
different spectral/morphological types
\citep*{Coleman_etal:1980}.
For each galaxy,  the SED (or linear combination of SEDs) that yielded
the observed  (F606W$-$F814W) colours at the galaxy's redshift was chosen
for the interpolation. The formal uncertainty in this transformation is very
small (${\rm rms}\sim0.01\,$mag), but that assumes perfect knowledge of the
filter response in each band.  Tests with different filters indicate that
systematic uncertainties of $\sim0.1\,$mag could be present, but these are
negligible in our analysis.
Absolute $B$ magnitudes corrected for internal extinction, $M^i_B$,
were calculated following \citet{Tully_Fouque:1985}. The correction, which
is a function of $i$, was in the range 0.32--0.96$\,$mag, with a median value
of 0.61$\,$mag.

The disk inclinations
($i$) were determined on the F814W images using the GIM2D programme
\citep{Simard_etal:2002}.
The uncertainties on $i$ were included in the
uncertainties on $\Vrot = (\Vrotsini) / (\sin i)$.
The inclinations were in the range 33--82$^\circ$
(90$^\circ$ being edge-on), with a median value of 68$^\circ$.

\section{Discussion of the Tully--Fisher relation}

The Tully--Fisher plot is shown in Fig.~\ref{fig:Tully-Fisher}(a).
Galaxies 1459 and F03 deviate strongly from the rest.
Galaxy 1459 may be two galaxies or a chain galaxy 
(cf.\ Fig.~\ref{fig:HST_montage}).
Galaxy F03 has a very low surface brightness (the lowest in the sample). 
These two galaxies will be excluded from the analysis.
In Fig.~\ref{fig:Tully-Fisher}(a) we plot the local Tully--Fisher
relation (TFR) from \citet{Pierce_Tully:1992}, as adapted by  
\citet{Vogt:1999:mn_quick_fix}.
Most of our field galaxies and all our cluster galaxies fall 
on the high luminosity/low velocity side of this relation.
The absolute magnitude residuals at fixed rotation velocity  
($\dTF$) are plotted against redshift
in Fig.~\ref{fig:Tully-Fisher}(b).
For the field galaxies, $\dTF$ becomes more negative with $z$
(94\% significance from a Kendall's tau test). 
Assuming that the errors in $\dTF$ are the individual measurement errors
plus 0.6$\,$mag added in quadrature
(a guessed intrinsic scatter in the \mbox{$B$--band}
TFR for this redshift range),
a chi-square fit for the field galaxies gives $\dTF = (-1.6 \pm 0.3) z$
with $\chi^2_{\rm r} = 1.0$.
If taken at face value and interpreted as luminosity evolution, 
the effect is $\sim\!0.8\,$mag at $z=0.5$, much larger than the
$\sim\!0.2\,$mag \citet{Vogt:1999:mn_quick_fix} found
at similar redshifts.
This interpretation assumes a non-evolving  TFR slope, which is  important
since our field sample has a built-in positive correlation between luminosity
and redshift.
Our sample does not allow us to constrain the TFR slope, but
\citet{Ziegler_etal:2002} found some evidence for a slope change at $z \sim
0.5$ for a sample of 60 field spirals. Thus, the derived luminosity evolution
of the field spirals must be regarded with extreme caution.
One might worry that the low $\dTF$ values instead reflected underestimated
rotation velocities due to not detecting the emission at sufficiently
large galactocentric distances.
In that case low $\dTF$ values should be accompanied by
low $\rdspec/\rdphot$ values,
and Kendall's tau tests for the cluster and the field galaxies
show this not to be the case.

In order to compare the cluster and field spirals,  we will use a constant TFR
slope equal to the local value. We will make  this comparison under two
alternative (and extreme) hypotheses. First, we will assume that the zero point
of the field TFR does not evolve with $z$ and compare the residuals from the
local TFR ($\dTF$) for the field and cluster spirals. Second, we will assume
that the zero point of the field TFR evolves with $z$ as
shown by the dotted line in Fig.~\ref{fig:Tully-Fisher}(b),
and compare the residuals from the
local TFR corrected for this evolution, $\dTFcorr \equiv \dTF - (-1.6 z)$.
In both cases we compare the cluster sample ($N_{\rm clus} \! = 7$)
with the full field sample ($N_{\rm field} \! = 18$) and a field subsample
spanning the luminosity range of the cluster galaxies ($M^i_B \! < \! -19.8$;
$N_{\rm field} \! = 7$). Two statistical tests are carried out,  a simple
difference of the mean values (assuming rms$/\sqrt{N}$ uncertainties) and a
Kolmogorov--Smirnov (K--S) test.  For comparison purposes, the K--S probability
that the field and cluster samples are drawn from  two different populations
has been translated into a number of sigmas for a normal distribution. 
The results of these comparisons are given in Table~\ref{tab:tests}. 
The difference between cluster and field samples are larger and
more significant using the first set of assumptions, but it is clear that  
the data suggest the cluster spirals are \mbox{$\sim\!0.5$--$1\,$mag}
brighter than
the field ones at a fixed rotation velocity.  
The only other published cluster TF study at intermediate $z$
is that of \citet*{Metevier_etal:2002:mn_quick_fix},
who found a larger TFR scatter for their 7 cluster spirals at $z=0.39$,
but no evidence for a zero point or slope change. The larger look-back-time
of our study could explain the fact that we do find some luminosity evolution. 

The increment in luminosity that we have found  could be the result of enhanced
star formation in spiral galaxies falling onto the cluster. However, these
suggestive results are far from conclusive since  we have studied one single
cluster and our current sample is small. It is clear that a similar  study of a
reasonable  cluster sample, spanning a broad redshift baseline,  can provide
strong empirical results on the evolution of the cluster spiral population. 

\begin{table}
\centering
\caption{Tully--Fisher differences: cluster versus field}
\begin{tabular}{llcr}
\hline
Variable   & Sample &
$\langle{\rm cluster}\rangle - \langle{\rm field}\rangle$ [mag] &
$\Pdiffdistr$ \\
\hline
$\dTF$     & All      & $-1.32\pm0.43$ ($3.1\sigma$) & 96\% ($2.0\sigma$) \\
$\dTF$     & High $L$ & $-1.00\pm0.46$ ($2.2\sigma$) & 87\% ($1.5\sigma$) \\
$\dTFcorr$ & All      & $-0.71\pm0.42$ ($1.7\sigma$) & 83\% ($1.4\sigma$) \\
$\dTFcorr$ & High $L$ & $-0.67\pm0.43$ ($1.6\sigma$) & 87\% ($1.5\sigma$) \\
\hline
\end{tabular}
\label{tab:tests}
\end{table}

\section*{Acknowledgements}

We wish to thank Luc Simard for providing the ELFIT2D and GIM2D
software, and for helpful discussions.
We are grateful to Pieter van Dokkum and \mbox{Marijn} Franx
for making their reduced WFPC2 images of {\MS1054} available.
Douglas Clowe is thanked for letting us use a deep
Keck image of {\MS1054}. 
Gianni Busarello is thanked for suggesting the method used to
remove the geometrical distortions from the spectra.
We thank the referee Nicole Vogt for constructive comments
that helped improve the presentation.
This work was supported in part 
by the Gemini Observatory, which is operated by the AURA
on behalf of the international Gemini partnership of
Argentina, Australia, Brazil, Canada, Chile, the UK, and USA\@.
Generous financial support from the Danish Research Training Council 
(BM-J) and
the Royal Society (AA-S) is acknowledged.

\bibliographystyle{mn2e}
\bibliography{papers_cited_by_milvang}

\label{lastpage}

\end{document}